\documentclass[twocolumn,10pt]{asme2ej}
\usepackage{graphicx}
\usepackage{enumerate}
\usepackage{amsmath,amssymb}  % Better maths support & more symbols
\usepackage{bm}  % Define \bm{} to use bold math1 fonts
\usepackage{pdfsync}  % enable tex source and pdf output synchronicity
\usepackage{placeins}
\usepackage{url}

\title{A Light-weight  Vibrational Motor Powered Recoil Robot that Hops 
Rapidly Across Granular Media}
\author{Alice C. Quillen
	\affiliation{\small Dept. of Physics and Astronomy, \\ University of Rochester, \\ Rochester, NY, 14627, USA\\
		  \texttt{alice.quillen@rochester.edu}}}
\author{Randal C. Nelson
	\affiliation{\small Dept. of Computer Science, \\ University of Rochester, \\ Rochester, NY, 14627, USA\\
		\texttt{nelson@cs.rochester.edu}}}
\author{Hesam Askari \affiliation{\small Dept. of Mechanical Engineering, \\ University of Rochester, \\ 
		Rochester, NY, 14627, USA\\ 		\texttt{askari@rochester.edu}}}
\author{Kathryn Chotkowski
	\affiliation{\small Dept. of Physics and Astronomy, \\ University of Rochester, \\ Rochester, NY, 14627, USA\\  
		\texttt{kchotkow@u.rochester.edu}}}
\author{Esteban Wright 
	\affiliation{\small Dept. of Physics and Astronomy, \\ University of Rochester, \\ Rochester, NY, 14627, USA\\ 
		\texttt{ewrig15@ur.rochester.edu}}}
\author{Jessica K. Shang	\affiliation{\small Dept. of Mechanical Engineering,\\	University of Rochester, \\ Rochester, NY, 14627, USA\\ 	\texttt{j.k.shang@rochester.edu}}}
\begin{document}
\maketitle

%\tableofcontents

%\footnote{
%\texttt{kchotkow@u.rochester.edu},
%\texttt{mfeehan2@u.rochester.edu},
%\texttt{enolting@u.rochester.edu},
%\texttt{askari@rochester.edu},
%\texttt{pmiklavc@u.rochester.edu},
%\texttt{scott\_seidman@urmc.rochester.edu},
%\texttt{j.k.shang@rochester.edu},
%\texttt{ruy.ibanez@gmail.com}
%}

\begin{abstract}
Abstract: {\it 
A 1 cm coin vibrational motor fixed to the center of a 4 cm square foam platform moves rapidly across  granular media (poppy seeds, millet, corn meal) at a speed of up to 30 cm/s, or about 5 body lengths/s. Fast speeds are achieved with dimensionless acceleration number, similar to a Froude number, up to 50,   allowing the light-weight 1.4 g mechanism to remain above the substrate, levitated and propelled by its kicks off the surface.   The mechanism is low cost and moves without any external moving parts. With 2 s exposures we photograph the trajectory of the mechanism using an LED  blocked except for a pin-hole and fixed to the mechanism.  Trajectories can exhibit period doubling phenomena similar to a ball bouncing on a vibrating table top.  A two dimensional numerical model gives similar trajectories, though a vertical drag force is required to keep the mechanism height low.  We attribute the vertical drag force to aerodynamic suction from air flow below the mechanism base 
and through the granular substrate. Our numerical model suggests that speed is maximized when the mechanism is prevented from jumping high off the surface. In this way the mechanism resembles  a galloping or jumping animal whose body remains nearly at the same height above the ground during its gait.
}
\end{abstract}

%{\bf Keywords:} zzzz

%target: Journal of Mechanisms and Robotics

%\url{https://journaltool.asme.org/home/JournalDescriptions.cfm?JournalID=23&Journal=JMR}
%\url{http://journaltool.asme.org/Help/AuthorHelp/WebHelp/JournalsHelp.htm}
%\url{https://www.asme.org/shop/journals/information-for-authors/journal-guidelines/writing-a-research-paper}

\section{Introduction}

Limbless locomotion by snakes or worms gives a paradigm for locomotion 
in rough and complex environments  \cite{childress81,hirose93,alexander03,guo08,childress12}.
Soft and hard robotic devices can crawl over a surface 
due to an asymmetric or directional  dynamic friction  (e.g., \cite{chernousko10,gidoni14,noselli14,wagner13,giomi13}).
Snakes and snake-like robots (e.g., \cite{hosoi15,maladen09,maladen11,hatton11}) propel themselves by exploiting asymmetry in the friction they generate on a substrate. 

Vibrating legged robots provide a different, but related example of locomotion that also
exploits frictional asymmetry (e.g., \cite{cicconofri15,scholz16}).   These are mechanisms
of minimal complexity that exploit periodic shape changes to propel themselves 
  (e.g., \cite{desimone12,giomi13,wagner13,noselli14}). 
Modulation of friction due to oscillations of the normal forces causes
stick-slip horizontal motions and net horizontal displacement. 
An example is the
 table top  bristlebot toy that can be constructed by fixing  a low-cost vibrational motor  to the head of a 
toothbrush \cite{bristlebot}.
%\footnote{\url{http://www.evilmadscientist.com/2007/bristlebot-a-tiny-directional-vibrobot/}}. 
Examples of vibrationally powered mechanisms include miniature robots 
(e.g., \cite{chatterjee05,jalili16,majewski17}) that are smaller than a few cm in length.

The granular medium presents additional challenges for a locomotor (e.g., \cite{hosoi15})
as propelled grains or particles can jam the  
mechanism, and exert both drag-like and hydrostatic-like forces \cite{tsimring05,katsuragi07,aguilar16,aguilar16b}.
A vibrating mechanism can sink into the medium causing mechanism to tilt or impeding its motion, 
or the mechanism may float due to the Brazil nut effect (e.g., \cite{knight93}).   
A variety of animals propel themselves rapidly
 across granular surfaces.  The lightweight zebra-tailed lizard  (10 cm long, 10g)
 moves 10 body lengths/s \cite{li12}.   The six legged  DynaRoACH  robot (10 cm, 25 g)  is a rotary
 rotary walker that 
approaches a similar speed (5 body lengths/s) on granular media \cite{zhang13}.

In this paper we work in the intersection of these fields, exploring
how light-weight, small, vibrating and legless locomotors a few cm in size can 
move rapidly on the surfaces of granular media.  
%such as grains (millet), seeds (poppy seeds), or sand.
An advantage of a legless locomotor is that appendages cannot get jammed or caught 
in the mechanism. 
As most small animals that traverse sand either use legs or wiggle,
our locomotors have no direct biological counterparts but they are similar to vibrating table top
toys.  They are in a class of mechanisms that locomote due to recoil from internal motions
(e.g., \cite{childress11}). 
Small vibrational motors are low cost, so if effective autonomous locomotors can be devised
with them,  large numbers of them could be simultaneously deployed for distributed exploration. 

In this manuscript we describe, in section \ref{sec:construct},
construction of a light weight (less than 2 g) and low-cost mechanism (a few dollars) that can rapidly
traverse granular media at a speed of a few body lengths per second.  We estimate
an acceleration parameter or Froude number for the mechanism, classifying it as a jumper
or hopper, in comparison to animal gaits.  
%Examination of the mechanism trajectory
%shows phenomena similar to that displayed by a hard ball bouncing on a vibrating table top.
Because the mechanism trajectories often show the mechanism touching the substrate once
per motor oscillation, we infer that there must be a force pulling the mechanism downwards.
In section \ref{sec:aero} we estimate that airflow through the granular medium could
affect the mechanism motion.  Rough measurements of air flow rate as a function of pressure
are used to estimate the size of this force in section \ref{sec:flow}.  
In section \ref{sec:model} we develop a simple 
two-dimensional model for the mechanism motion.

\section{Mechanism Construction and Experimental Setup}
\label{sec:construct}

\begin{figure}
\centering
\centerline{
\includegraphics[width=3.8in]{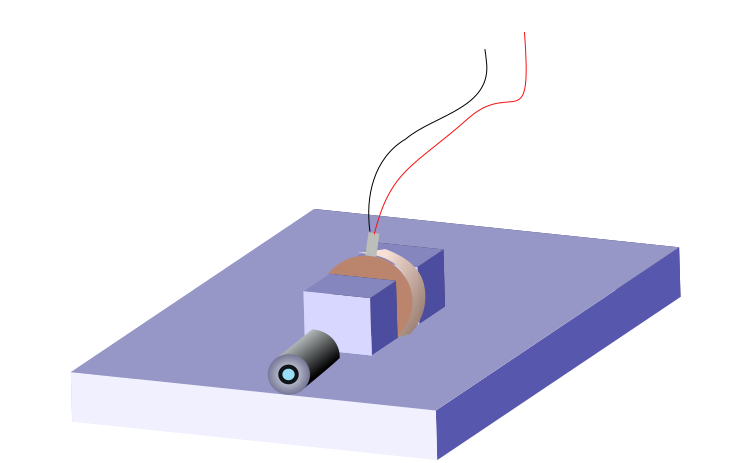} }
\caption{An illustration of the hopper mechanism.   A 1 cm diameter vibrational motor is fixed to a  
4 $\times$ 4  $\times$ 0.5 cm foam platform.  
An LED is used to track its motion.
  The LED is covered with foil tape that has
been punctured with a pinhole. Wires and a resistor for the LED are not shown.
\label{fig:mechanism}}
\end{figure}

We place a  5VDC coin vibrational motor on a light rigid foam platform (see Fig. \ref{fig:mechanism}).
The motor is 1 cm in diameter and rotates at approximately 12000 rpm.
The foam platform is rectangular and a few cm long but only a few mm thick.
The foam is closed cell, moisture-resistant rigid foam board, comprised of
extruded polystyrene insulation (XPS)  
made by Owens Corning (with brand name Foamular Pink). 
Its density is apparently 1.3 pounds per cubic foot which is 0.0208 g/cm$^3$.
The coin vibrational motor is oriented so a flat face is perpendicular to the foam
board.  To rigidly fix the motor to the foam board platform we used double sided tape
attached on either side of the motor and to small foam blocks which are then 
attached to the platform, as shown in Fig. \ref{fig:mechanism}.
The motor is externally powered with a DC power supply and via light and ultra flexible wire so as not to interfere 
with motion. 
The wire we used is a thin and flexible multi-strand silver plated copper
36AWG wire with silicone rubber insulation.  %Calmont Siliflex brand.

\begin{figure}
\centering
\centerline{
\includegraphics[width=2.0in]{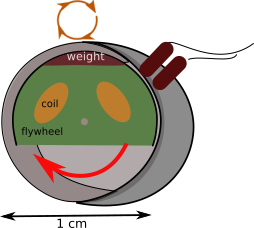} }
\caption{An illustration of the inside of a coin vibrational motor.  
The motor case is rigid. A lopsided flywheel inside rotates
causing the entire case to move in a circle in a direction countering the motion of the internal weight.  
If the mechanism is prevented from rotating and
in the absence of external forces,
each point on the surface of the case moves in a circle, here shown with a brown circle with arrows.  
%The case surface  translates first downwards, then to the left, then upwards, and then to the right (or vice versa
%if the flywheel inside the motor rotates in the opposite direction).
\label{fig:motor}}
\end{figure}

Inside the vibrational motor is a lopsided flywheel, giving a displacement between
the center of mass of the motor and its case.   When the motor rotates, recoil from the flywheel
causes the motor case to vibrate back and forth and up and down.  
In the absence of external forces, and if prevented from rotating, the motor
case moves in a circle, as illustrated in Fig. \ref{fig:motor}.   The motor is designed
to rotate at 12,000 rpm (200 Hz) at 5VDC. However, we have found that the frequency depends
on voltage, ranging from $f \approx 180$  to 280 Hz over a voltage range of 3.5 to 5.5V.

To track motor motion, we attached a small %(3 mm size) 
clear blue LED to the platform, 
oriented perpendicular to a flat
motor face.  The LED is powered in series with a 100 $\Omega$ resistor.  
The LED is covered with foil
tape that has been punctured by a pinhole and is powered with the same DC power lines as used
to power the motor.   The mechanism weight  totaled 1.4 g.   

\begin{figure*}
\centering
\centerline{
\includegraphics[width=6.0in]{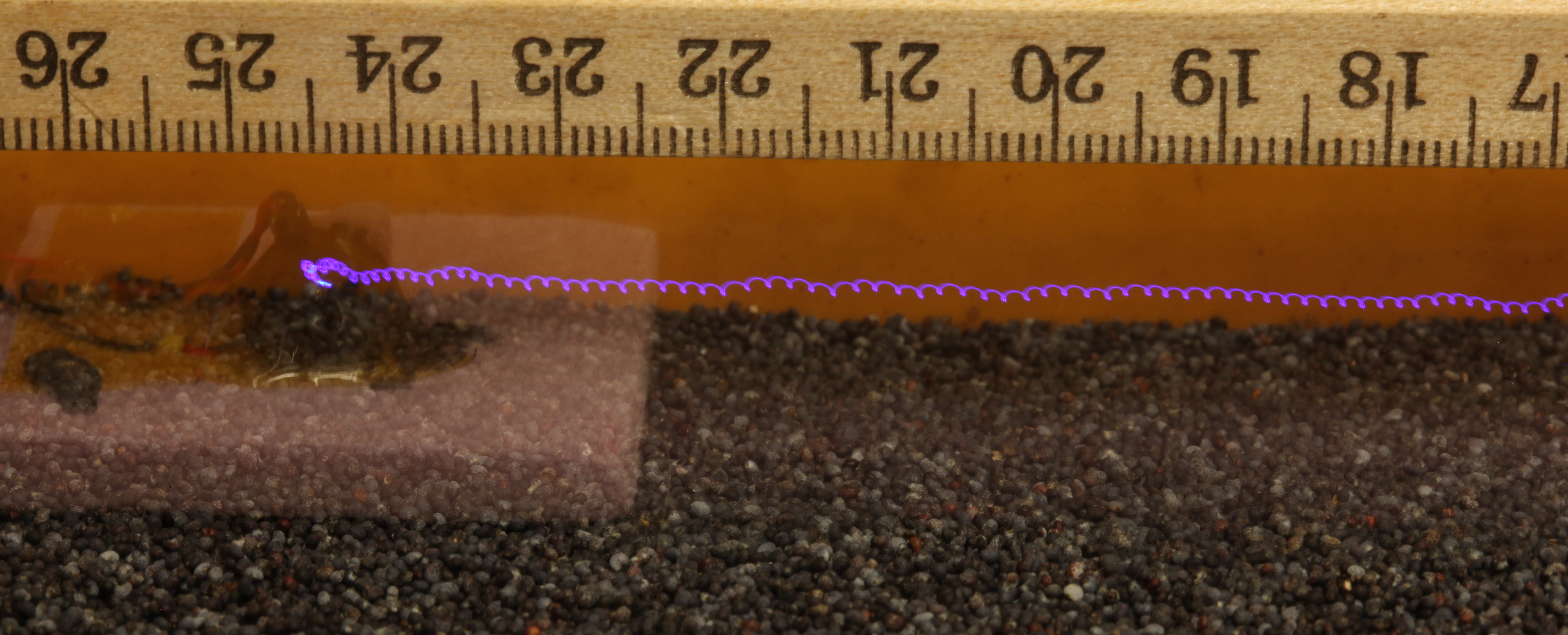} }
\caption{Photograph showing a 2 s exposure in ambient room lighting. 
The exposure was started when the mechanism was stationary and on the left.
Then the motor was turned on, and the mechanism moved to the right.
A blue LED is mounted on the mechanism.  
The LED is covered with foil tape punctured with by a pin hole.  The bouncing trajectory of
the mechanism is traced by the moving position of the pinhole, giving  a series of loops as the 
mechanism hops to the right.  We measure the speed of the mechanism by counting the loops,
 using the frequency of the motor and distances measured with the ruler mounted above the mechanism.
The substrate in this photograph is black poppy seeds and the motor frequency is 220Hz.
%Subsequent photographs were taken in darkness.
\label{fig:setup}}
\end{figure*}

Our experimental setup is shown in Fig. \ref{fig:setup}.  The mechanism is placed
on a flat granular bed. The grains are poppy seeds, cornmeal or millet.   We don't press or compactify
the medium but do sweep it flat prior to taking photographs.
Granular materials can be described by an angle of repose $\theta_{\rm repose}$.
By tilting the trays holding the media, 
 we measured  $\theta_{\rm repose} \approx 34^\circ$ for the 
 granular materials used here;  poppy seeds, cornmeal and millet.  
The angle of repose is related to the coefficient 
of static friction, $ \mu_s$, with  
$ \tan \theta_{\rm repose} = \mu_s$.
For our granular materials the coefficient of static friction $\mu_s \sim 0.7$.

\begin{figure*}
\centering
\includegraphics[width=6.0in, trim={20mm 40mm 0mm 10mm},clip]{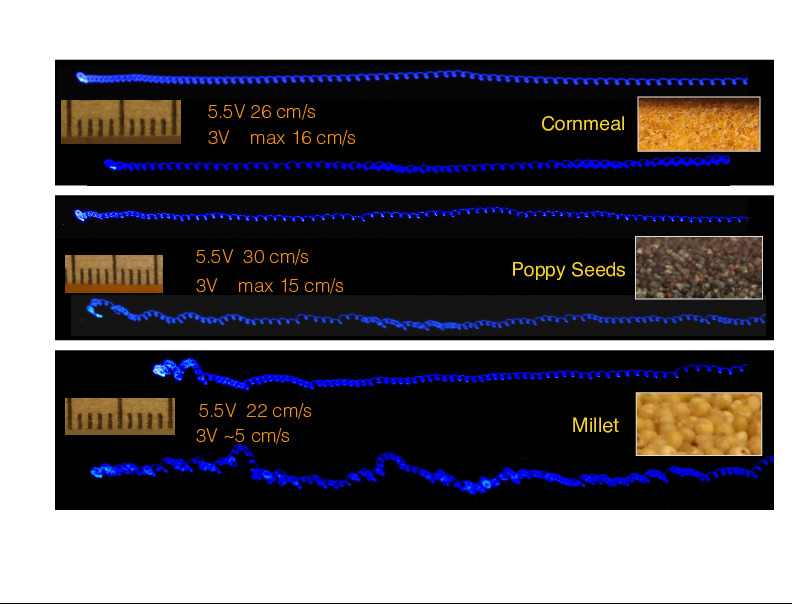} 
\caption{Mechanism trajectories on three different substrates and at two different motor voltages.
The blue lines show the tracks of the blue LED during 2 second exposures.
These show the motion of the mechanism
as it traversed the granular medium.
The top pair of trajectories show the mechanism on cornmeal, the middle pair on poppy seeds and the
bottom pair on millet.   Sub-panels show a 1 cm scale and views of the granular substrate.  The subpanels
were extracted from the setup photographs without rescaling them.  These
were taken prior to the long exposure photographs and with the same camera setup.
At 3V the motor frequency was 175 Hz whereas at 5.5V it was 285 Hz.
Mechanism horizontal speeds are estimated from the number of loops travelled per cm.
Mechanism displacement $A$ caused by motor recoil is estimated from the vertical
amplitude of the trajectory loops.
\label{fig:wiggle}}
\end{figure*}

\begin{figure*}
\centering
\includegraphics[width=5.0in]{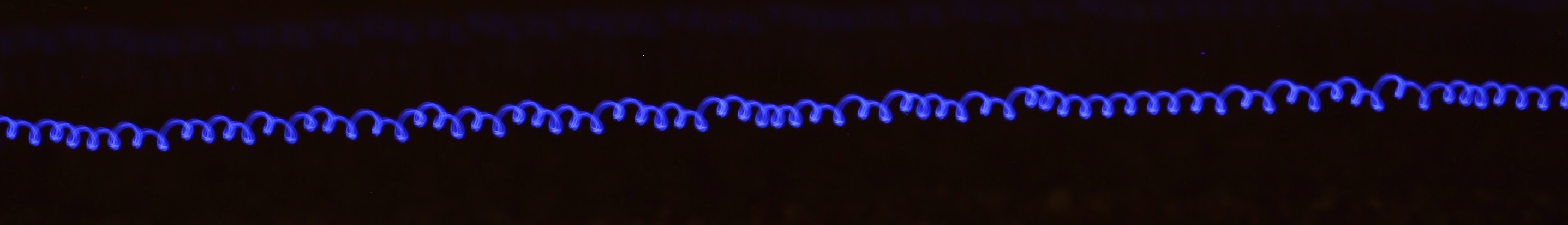} 
\caption{A photographed trajectory illustrating period doubling and tripling. 
This a trajectory of a somewhat more massive mechanism (1.7g) with a 
larger displacement ($A\sim 0.5$ mm), a frequency of 240Hz and on cornmeal.
\label{fig:doubling}}
\end{figure*}

A camera with a macro lens was
used to photograph the mechanism in motion.  We open the camera shutter and then turn on the motor.
With a 2 second exposure, the motion of the mechanism is tracked with the light from the LED,
as shown in Fig. \ref{fig:setup}.   With the room lights on we can see the original location
of the mechanism and the LED track as the motor moved to the right.    Subsequent photographs,
shown in Fig. \ref{fig:wiggle} and Fig. \ref{fig:doubling},
were taken in the dark and show only the tracks made by the LED.    
A ruler mounted above the mechanism (see Fig. \ref{fig:setup}) gives the scale in cm. 
We have also filmed the motion of the mechanism with a high speed camera at 1000 frames per second
(see Videos 1 and 2).

\begin{figure*}
\centering
\includegraphics[width=2.0in]{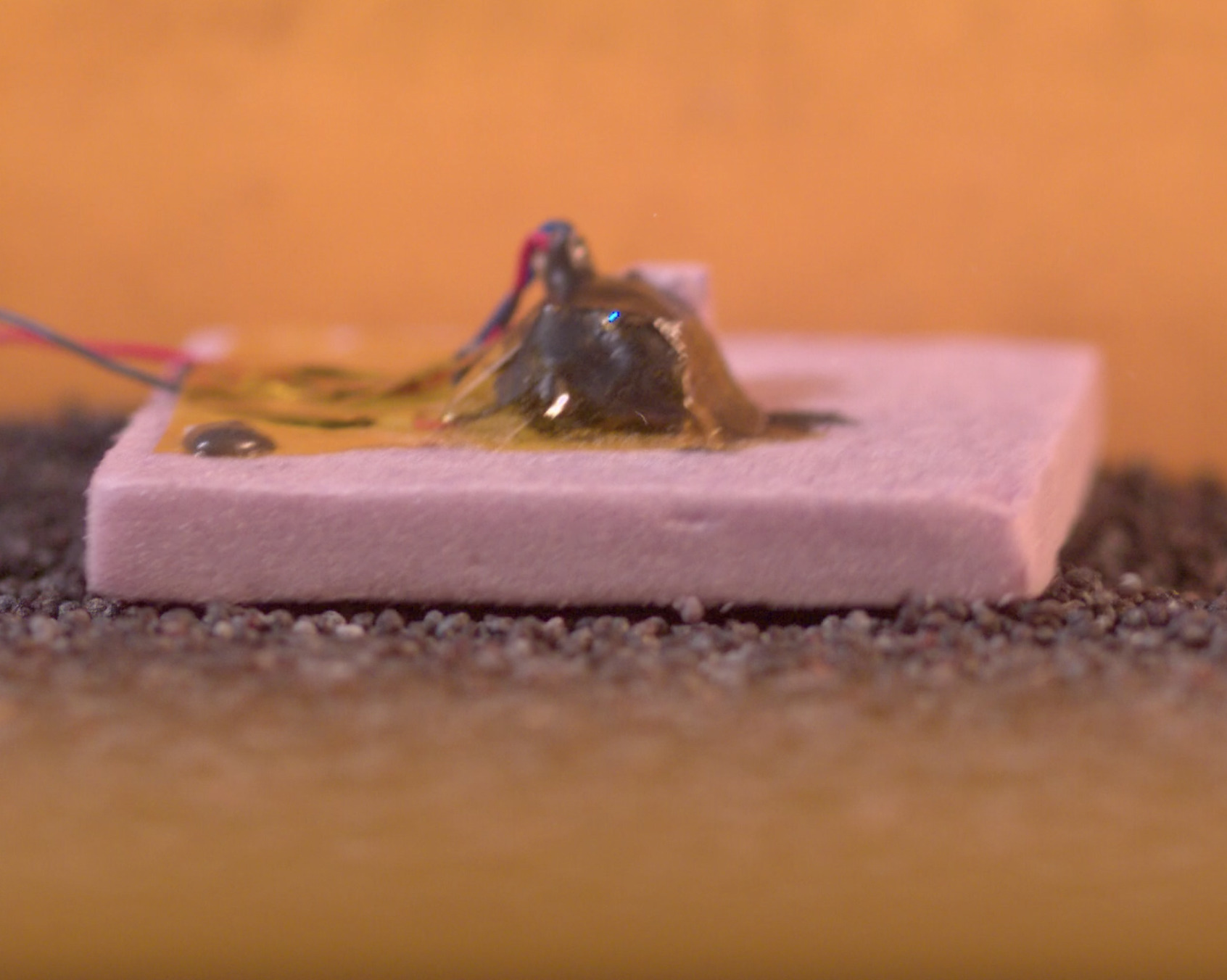} 
\includegraphics[width=2.0in]{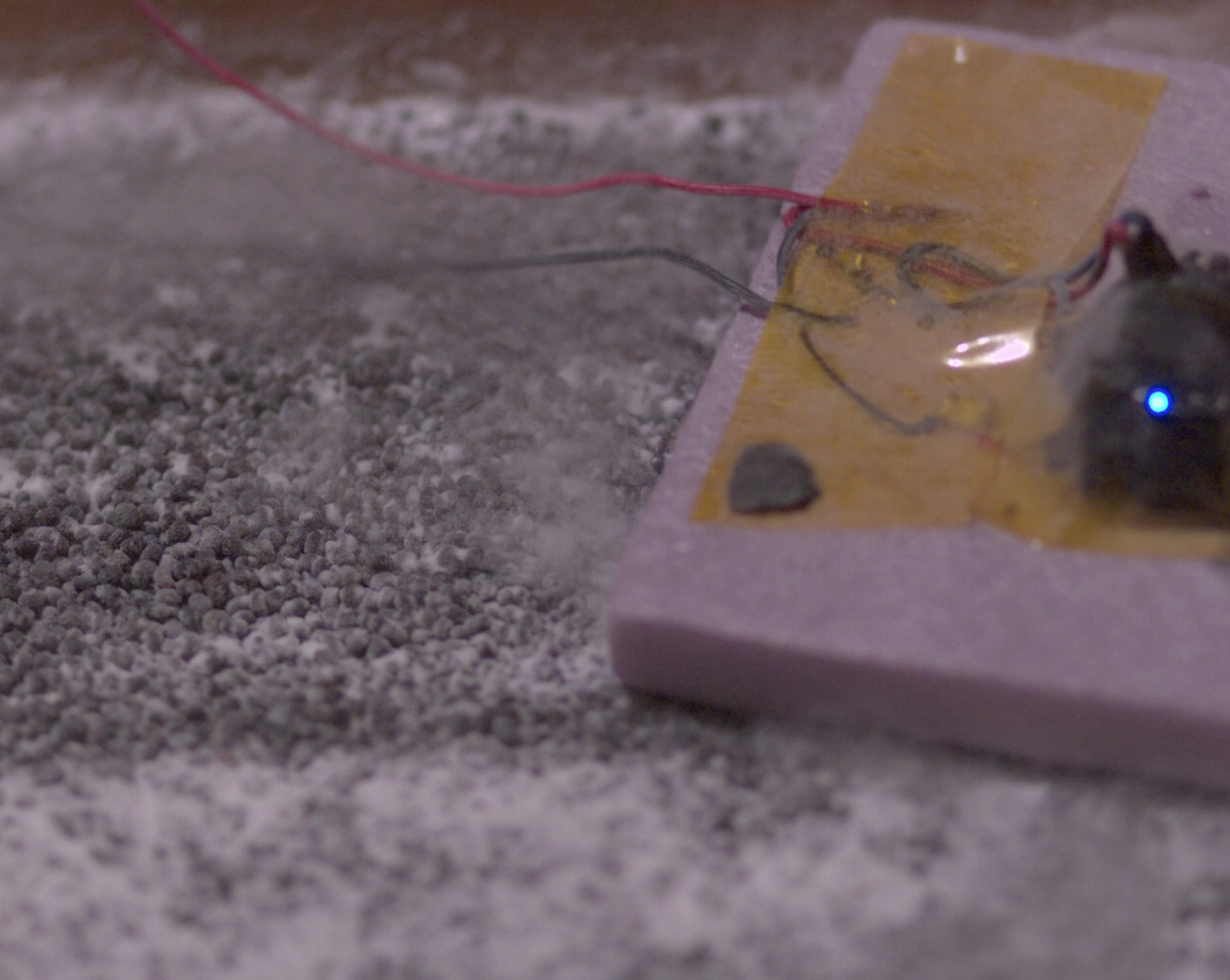} 
\caption{Place holder for two high speed videos.  Both show the same mechanism traversing
poppy seeds and filmed at 1000 frames per second.  The poppy seed substrate has been covered
with a layer of cornstarch in the second video. 
Videos are temporarily viewable here \protect\url{https://youtu.be/kLsG6e_-IqM}.  
\label{fig:videos}
}
\end{figure*}

We took audio recordings
of the mechanism while in motion.  We measured the dominant frequency present in the sound files 
and this gave a measurement for the vibrational motor frequency.   This way we could determine the frequency
of vibration at different DC voltages and for specific motors. 
For the trajectories shown in Fig. \ref{fig:wiggle}, 
the audio recorded motor frequencies were 175,  220, and 285 Hz at 3, 4 and 5.5V, respectively.
With the motor frequency,
and by counting the number of loops per centimeter in the LED trajectory, we can compute the
speed of the mechanism as it traverses the granular medium.
Fig. \ref{fig:wiggle} shows three pairs of photographs showing the mechanism moving across 
 three different substrates, cornmeal, poppy seeds and millet.
On each substrate we set the motor voltage to 3V or 5.5V.
The lower trajectory in each pair shows the lower voltage setting.
Measured horizontal speeds are labelled as text on Fig. \ref{fig:wiggle} for each trajectory.
For each experiment
  we took a photograph of the experimental setup with the same camera setup and focus.
An extracted region of the ruler from the setup photographs gives a scale and
 are shown in the sub-panels in Fig. \ref{fig:wiggle}.
 From the same setup photographs, we also cut out a sub-panel  showing the granular medium. 

Mechanism trajectories seen in Fig. \ref{fig:wiggle} exhibit regions of periodic behavior.
During these times the mechanism touches the substrate once per motor oscillation period.
Period doubling  describes when
the mechanism touches down once for every two   oscillation periods. 
The figure also shows regions where period doubling  occurs; for example, see the lower trajectory
on cornmeal.    Fig. \ref{fig:doubling} shows a trajectory from a heavier mechanism and a motor
giving a larger displacement at 240 Hz on cornmeal that illustrated period doubling and tripling.

To measure the motor recoil we filmed a bare vibrational motor
hanging from a thread.  Horizontal  peak to peak motions were about 1 mm 
giving an amplitude of vibrational motion of about $A_{\rm bare} \sim 0.5$ mm.
The motor itself weighs only $m_{\rm motor} \sim 0.9$ g whereas the entire mechanism 
is more massive, $M \sim 1.4$ g.
With the motor affixed to the platform, the amplitude of motion for the entire
mechanism (in free space) depends on the ratio of the mass of bare motor $m_{\rm motor}$ to mass
of mechanism $M$,
\begin{equation}
 A \sim A_{\rm bare} \frac{m_{\rm motor}}{M} \sim 0.3\ {\rm mm}. 
 \end{equation}
We can also estimate the recoil amplitude from our photographs.  Amplitude $A \sim 0.3$ mm
is consistent with the vertical length of the loops in the trajectories of Fig. \ref{fig:wiggle}.

\begin{table}
\centering
\caption{Approximate Quantities \label{tab:quantities}}
\begin{tabular}{lll}
\hline
%$\omega$  & vibrational motor angular frequency & $\sim 1200 {\rm s}^{-1}$ \\
$f$   & vibrational motor frequency & 200 Hz  \\
        &                                              & (12000 rpm)\\
$\omega$  &  motor angular frequency & $1257 {\rm s}^{-1}$ \\
$m_{\rm motor}$   & mass of vibrational motor & 0.9 g \\
%$m_r$  & recoil mass \\
$M$   & mass of mechanism & $1.4$ g \\
$L$  & dimension of platform &  $ 4$ cm\\
$A $ & amplitude of motion   & $  0.3 $ mm \\
         & (whole mechanism) & \\
$2A_{\rm bare}$ & displacement of bare motor & $ 1.0$ mm  \\
 & ( peak to peak) &   \\
$A \omega $ & velocity of vibration   & $  38 $ cm/s \\
$A \omega^2 $ & acceleration of vibration  & $475 $ m s$^{-2}$\\
\hline
$g$  & gravitational acceleration & 9.8 m/s$^2$\\
%\hline
$\rho_g$  & density of granular medium & $ 1$ g/cc \\
%$m_s$ & mass of a single grain & \\
%$s$  & size of grains & 0.4 -- 2 mm \\
%$\theta_{repose}$ & critical angle of repose & $\approx 34^\circ$\\
$\mu_{s}$ & coefficient of static friction & $\approx 0.7$ \\
$\rho_{\rm air}$  & density of air & $1.2\times 10^{-3}$ g/cc \\
$\nu_{\rm air}$   & kinematic viscosity of air & $ 1.5\times10^{-5}\ {\rm m}^2 {\rm s}^{-1} $\\
$\mu_{\rm air}$   & dynamic viscosity of air & $ 1.8 \times10^{-5}\ {\rm Pa~s} $\\
\hline
\end{tabular}\\
%Notes: I measured $2A_{bare}$ from a video, but it would be better measured from the photographs. 
%\section{Grain sizes}
Grain size diameters were measured with a caliper, giving
$d=0.62$ mm for poppy seeds.
$d=0.45$ mm for cornmeal and 
$d=1.7-2.2$ mm millet.  The range is given because they are not spherical.
%LD polyethyline platform: density is 0.0157 g/cc. siliflex wire is  0.00292 g/cm.
\end{table}

When turned on, 
a vibrational motor sitting on the surface of a granular medium  digs a small crater and remains
vibrating in the bottom of it,
rather than moving across the surface.  The foam platform of our mechanism distributes the  force
on the granular substrate 
associated with the motor motion.  For these mechanisms the granular medium is barely disturbed
as the mechanism moves across it.   Only on cornmeal is a faint track left behind as the
mechanism moves across it.  % and this is probably due to electrostatic interactions.  
On millet and poppy seeds,  before and after
photographs showed that only a few grains were disturbed after the mechanism traversed the surface.   
Granular media is often described in terms of a flow threshold or critical yield stress \cite{bagnold54}.   
At stresses below the critical one, grains do not move.
Our hopper mechanism exerts such small pressure onto the granular medium that
the critical yield stress of the granular medium is not exceeded.   
There are trade-offs in choosing the surface area and thickness of the foam platform.  If the mechanism is too heavy,
it won't jump off the surface and its speed is reduced.  If the platform is too small, then the vibrating
mechanism craters instead of moving across the surface.  If the platform is too thin, it flexes and this
can prevent locomotion if the corners vibrate and dig into the medium.  Smaller platforms are less stable
as irregularities in the substrate can tip them.  The fastest mechanisms have stiff platforms, and motors
mounted low and centered on the platform.  Wires and LED are best taped to the platform
so that they  don't vibrate while the mechanism is moving.

The acceleration of the mechanism base
\begin{equation}
A \omega^2 = 475\ {\rm m~s}^{-2} \left( \frac{A}{0.3\ {\rm mm}} \right)
\left( \frac{12000\ {\rm rpm}}{f} \right)^2 
\end{equation}
where  
$\omega = f/(2\pi)$ is the angular frequency of vibration.   
%Agrees with Randall and with the observation that the motor alone on a table top can kick itself off the table.
The vibration causes the mechanism  to move with  velocity,
\begin{equation}
 A \omega = 38\ {\rm cm~s}^{-1}  \left( \frac{A}{0.3\ {\rm mm}} \right)
\left( \frac{12000\ {\rm rpm}}{f} \right) .
\end{equation}
This is a maximum velocity for the mechanism's horizontal motion.
The horizontal velocities reached by our mechanisms can be up to 0.6 times the maximum.

\subsection{Acceleration parameter or gait Froude number}
\label{sec:froude}

We estimate the time it takes the mechanism to fall the distance of the vibration or displacement
amplitude
\begin{equation}
 t_g = \sqrt{\frac{A}{g}},
 \end{equation}
 where $g$ is the acceleration due to gravity.
 % and $A$ is displacement about center of mass if in free space.
We can derive a  dimensionless quantity by comparing this time to the vibration angular frequency,
 \begin{equation}
t_g \omega=  \sqrt{\frac{A}{g}}  \omega .
 \end{equation}
The ratio of the   acceleration due to vibrational oscillation $A \omega^2 $
and that due to gravity $g$ is
\begin{equation} \Gamma = {\bf Fr} \equiv \frac{ \omega^2 A}{g},  \label{eqn:Froude}
\end{equation}
and this is equivalent to  $(t_g \omega)^2$.
This dimensionless number
 is also the ratio of centripetal force to gravity force and is also known as a walking or gait Froude number.
Gait frequency for animals scale with Froude number \cite{alexander84}, with a walker having ${\bf Fr} \lesssim 1$.

The dimensionless ratio $\Gamma$  is equivalent to an acceleration parameter  
used to classify the dynamical behavior of a hard elastic object bouncing 
on a vibrating plate but computed using the displacement and frequency of the table
rather than the mechanism  (e.g., \cite{holmes82,bapat86,germinard03,altshuler13,chastaing15}).
Using the amplitude of motion and  motor frequency of 280 Hz, we estimate
an acceleration parameter for our mechanism
of about $\Gamma \sim 48$.  
As $\Gamma \gg 1$,   our mechanism can be considered a hopper or a galloper rather
than a walker.   Gaits in animals depend on Froude number with
the transition to galloping taking place at about ${\bf Fr} \sim 4$ \cite{alexander83,alexander84}.

Our mechanism can launch itself off the surface with a velocity $v_0 = A\omega$
so the mechanism base should reach a maximum height above the substrate of  
\begin{align}
h_{\rm max} &= \frac{v_0^2}{2g}  =  \frac{\Gamma A}{2}.
\end{align}
It should remain airborne for a time
\begin{align}
t_{\rm airborne} = \frac{2A\omega}{g} = \frac{2 \Gamma}{ \omega} = \frac{\Gamma}{\pi} P_{\rm osc}, \label{eqn:airborne}
\end{align}
with $P_{\rm osc} = 2 \pi/\omega$ the oscillation period.
The acceleration parameter sets the maximum height reached by the mechanism base.
For an acceleration parameter of about 36 (for the 3V trajectories in Fig. \ref{fig:wiggle}) 
the height reached by the mechanism should be $15A$
and it should remain airborne for about 9 oscillation periods.
%The size of the displacement amplitudes can be seen in Figs. \ref{fig:wiggle}.
The maximum height reached during a long jump by the mechanism is approximately equal to $A\Gamma/2$ 
(for $A \sim 0.3$ mm, $\Gamma\sim 36$, $h_{\rm max} \sim 5$ mm) 
and that is consistent with the maximum heights sometimes reached in the 3V trajectories
shown in Fig. \ref{fig:wiggle}.
However, the photographed trajectories show that most of time  the mechanism touches the 
surface once per motor oscillation period and remains
within $2 A$ from the surface.     There must be an additional force preventing the mechanism from
leaving the surface at a velocity $A \omega$.  
If the medium allowed the mechanism to bounce, then the maximum height could be even higher.
Friction cannot pull the mechanism downward and
a downward force is needed to keep the mechanism from reaching larger heights  than
observed.   

We consider possible additional forces that could reduce the mechanism's upward vertical velocity and jump height.
We have included two supplemental  high speed videos taken at 1000 frames per second (fps)
 of the mechanism moving across
poppy seeds.  In the second video, the poppy seed substrate was covered with a light layer of cornstarch
prior to recording.
In the high speed videos, we did not see the platform rock or flex
much,  though waves excited by vibration can propagate down the power wires.
% and the wires connecting the LED can vibrate.    
It is unlikely that these two types of motion could consistently pull the mechanism downward as stiffer mechanisms
with a polystyrene platform base
behaved similarly to those with softer platform bases made of polyethylene foam. 
This led us to consider the role of aerodynamics.
We found that a mechanism with a flat platform base on a very flat surface (a glass sheet) horizontally 
moved slowly, but
after we poked holes in the platform, its speed across the surface was increased.   
We found that a mechanism on a solid plate 
containing holes jumped higher and more irregularly than when moving on a granular surface
or a solid flat plate.  A mechanism under vacuum  (1/100-th of an atmosphere) displayed more irregular motion
than the same mechanism under atmospheric pressure.
%We saw no evidence of compactification after the mechanism traversed granular media.
A mechanism moving over a granular medium covered in a light powder (cornstarch), see our second
supplemental video, blew the powder
an inch away from the mechanism as the mechanism moved across the medium.
These experiments imply that the air flow beneath the mechanism affects its motion.

\FloatBarrier

\section{Aerodynamics}
\label{sec:aero}

We estimate the pressure that would develop under
the platform using Bernoulli's principle.
We estimate the pressure with $\rho_{\rm air} v^2$ where $\rho_{air}$ is the density of air and $u_{\rm air}$
is the horizontal velocity of air under the mechanism platform.  
Placing the pressure in units of acceleration on the mechanism we estimate an acceleration on the mechanism
 \begin{equation}
 a_B \sim \frac{\rho_{air} u_{air}^2 L^2}{M},
 \end{equation}
where $L^2$ is the surface area of the mechanism platform base.
In units of the acceleration $A \omega^2$ 
the acceleration estimated with Bernoulli's principle
\begin{align}
\frac{a_B}{A \omega^2} &\approx \left( \frac{u_{\rm air}}{A \omega} \right)^2 \frac{\rho_{air} A L^2}{M} \nonumber \\
 &\approx 0.007   \left( \frac{u_{\rm air}}{A \omega} \right)^2 
 \left(\frac{L}{4 {\rm cm}} \right)^2
  \left(\frac{\rho_{\rm air}}{ 1.2 \times 10^{-3}\ {\rm g/cc} }\right) \nonumber \\
  & ~~ \times
    \left(\frac{ 1.4\ {\rm g}}{M} \right)
    \left(\frac{A}{ 0.3\ {\rm mm}} \right).  \label{eqn:Bernoulli}
\end{align}
The ratio is low, suggesting that aerodynamics cannot affect the mechanism dynamics.

The horizontal air velocity exceeds the vertical mechanism velocity.
We relate the volume flow rate of air between mechanism and substrate  
to the vertical velocity of the mechanism platform,
 $v_z$,  the  platform's surface area and $z$,  its height above the substrate;
\begin{equation}
\frac{dV}{dt} \approx 4 L z u_{\rm air} = L^2 v_z . \label{eqn:dVdt}
\end{equation}
This gives
\begin{align}
\frac{u_{\rm air}}{A \omega} & \approx \frac{L}{4z} \frac{v_z}{A\omega }  \nonumber \\
&\approx 30 
\left(\frac{L}{4\ {\rm cm}} \right)
\left(\frac{A}{z } \right)
\left(\frac{0.3\ {\rm mm}}{A} \right)
\left( \frac{v_z}{A\omega }   \right)
\end{align}
If we insert this into Eqn. (\ref{eqn:Bernoulli}), we find that
the estimated  pressure force would only be
significant when the platform is quite near surface, $z \sim A$,   and in that setting we should  
consider flow through the narrow space between platform and substrate and through the substrate itself.

The kinematic viscosity of air $\nu_{\rm air}$ in units of the
acceleration parameter $A^2\omega$ is not that 
small,
\begin{equation}
\frac{\nu_{\rm air}}{A^2\omega} \approx 0.1 \left(\frac{\nu_{\rm air}}{0.15\ {\rm cm}^2\ {\rm s}^{-1} }\right) 
\left( \frac{0.3\ {\rm mm}}{A}\right)^{2} 
\left(\frac{1257\ {\rm s}^{-1}}{\omega} \right).
\end{equation}
As our grain sizes are similar to the amplitude $A$, air motions 
close to or in between grains may be in the low Reynolds
number regime.

\begin{figure}
\centering
\includegraphics[width=3.5in]{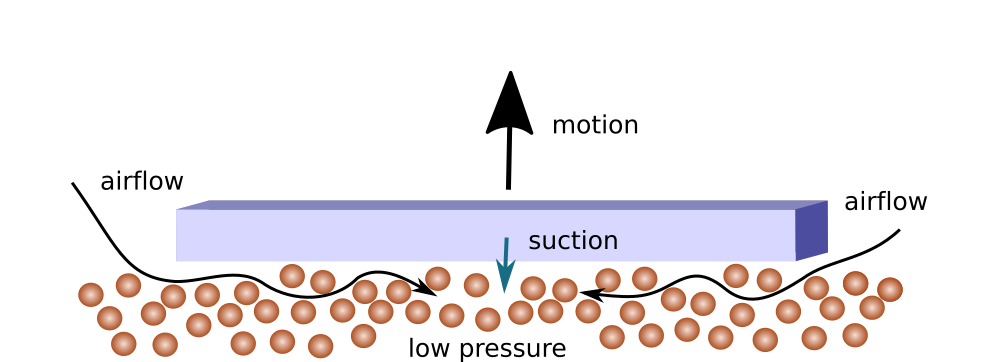} 
\caption{For the mechanism to launch itself off the substrate, air must flow beneath its platform base.
The air  pressure  must be lower than the ambient pressure under the platform for the air to flow
under the mechanism platform.  The lower pressure exerts suction on the platform base.
The permeability of the space between platform and substrate and of the
granular substrate could affect the mechanism locomotion.
\label{fig:permbase}}
\end{figure}

We consider the mechanism at rest after landing on the  substrate.
When the mechanism is pushing off
the surface, the air speed between grains would be low.  Because
the Reynolds number is proportional to speed,  the Reynolds number could be low
and so air viscosity could be important. 
Because of irregularity in the surface we can consider the gaps between
the mechanism platform and the granular medium as a permeable medium that air
must flow through.  See Fig. \ref{fig:permbase} for an illustration.

In a permeable or porous medium
Darcy's law relates the flow velocity $\bf u$  in a fluid to the pressure gradient ${\boldsymbol \nabla} p$
\begin{equation}
{\bf u} = -\frac{\kappa}{\mu} {\boldsymbol \nabla} p \label{eqn:Darcy}
\end{equation}
with $\mu = \nu \rho$ the dynamic viscosity and $\kappa$  the porous medium's permeability.
Permeability has units of area and depends on the interstitial spaces or the pore sizes.
 % it is similar to
 %the square of a typical pore diameter $\kappa \sim d_{pore}^2$ for interstitial spaces in the medium. 
%For loose sand the permeability ranges from $\kappa \sim 10^{-5} -10^{-8} {\rm cm}^{2}$.
An air pressure gradient is required for air to flow beneath the mechanism base.  For air to flow
under the mechanism, the pressure must be lower under it than outside it.   This causes
a suction on the mechanism.   As the air flow velocity $u \propto v_z$ is approximately proportional to the 
vertical velocity  of the platform, 
Eqn. (\ref{eqn:Darcy}) suggests that the force on the mechanism (or pressure per unit area)  due to air pressure
is proportional to $v_z$.  This gives a damping force on the mechanism that is important only when the
mechanism is touching or nearly touching the granular medium.
Furthermore air suction can pull the mechanism down, preventing it from reaching the
height we predicted in section \ref{sec:froude} using the acceleration parameter.

\subsection{Half-space flow field in a permeable medium}
\label{sec:half}

To estimate the size of a force due to air pressure on the mechanism base, we consider a 
circular platform with radius $r_h = L/2$ on a permeable medium.
We describe the air pressure and flow velocity in the permeable medium
in cylindrical coordinates 
$r,z$ and assuming azimuthal symmetry.    
Here $z<0$ below the granular substrate, $z=0$ on the  surface, and the origin is in 
the center of the platform that is touching the granular medium but at the moment it lands or takes
off vertically from the surface. 
The air pressure on the surface  is $p(r,z=0)$.  We take atmospheric pressure
to be zero (describing pressure with a difference from atmospheric) 
so $p(r_h,0)=0$ on the edge of the platform.  

We use Darcy's law  (Eqn. \ref{eqn:Darcy}) to relate air flow velocity ${\bf u}$ to the air pressure gradient
in the granular medium. 
Darcy's law combined with the condition for incompressible flow, ${\boldsymbol \nabla} \cdot {\bf u} = 0$,
yields Laplace's equation for pressure ${\boldsymbol \nabla}^2 p =0$.  
The boundary conditions determine the solution for the pressure $p(r,z)$. The flow field
is then set by the pressure gradient.

On the half space with $z<0$ a solution to Laplace's equation in cylindrical coordinates
that is well behaved at the origin and large negative $z$ is 
\begin{equation} 
p_k(r,z) \propto J_0(kr) e^{kz} ,
\end{equation}
where $J_0$ is a Bessel function of the first kind.
The general solution would be a sum or integral (over $k$) of such terms.
Because the pressure is zero at $r=r_h, z=0$, the Bessel function must have a root at $r=r_h$.
The first root of $J_0(x)$ is at $x\approx 2.4$.
We approximate the pressure profile under the platform with a single Bessel function
\begin{equation}
p(r,z) = p_0  J_0( 2.4 r/r_h) e^{ 2.4 z/r_h}. \label{eqn:prz}
\end{equation}
where the peak pressure under the mechanism is $p_0$.

We take the derivative of this solution with respect to $z$ and use Darcy's equation to estimate the 
vertical component of the flow velocity
\begin{align}
{u}_z(r,0) &= - \left. \frac{\kappa}{\mu_{\rm air}} \frac{dp}{dz} (r,z) \right|_{z=0}  \nonumber \\
&= u_c J_0(2.4r /r_h)   \label{eqn:uz}
\end{align}
with central velocity
\begin{equation}
u_0 =- \frac{\kappa}{\mu_{\rm air}} \frac{2.4}{r_h} p_0. \label{eqn:u0}
\end{equation}
It is convenient to compute $\int_0^{2.4} dx\ x J_0(x) = 1.24$.
The average $\bar u_z = \frac{1}{\pi r_h^2} 2 \pi \int_0^{r_h}r \ dr \ u_z(r,0) \sim u_0/2$.
The pressure integrated over the base area $2 \pi \int_0^{r_h} p(r,0) r\  dr \sim 1.4 p_0 r_h^2$.

With the platform vertical velocity $v_z$ equal to the average  $\bar u_z$,
we estimate the force on the mechanism from  pressure integrated over
the area of the platform base and using Eqn. (\ref{eqn:uz}) 
\begin{align}
{F}_{\rm aero} &\sim  2 \pi \int_0^{r_h} p(r,0)r\  dr   \sim 1.4 p_0 r_h^2 \nonumber \\
&\sim -\frac{\mu_{\rm air}}{\kappa} {r_h^3} v_z \sim -\frac{\mu_{\rm air}}{\kappa} \frac{L^3}{8} v_z.
\label{eqn:Faero}
\end{align}
%\int x J_0(x) from 0 to 2.4 is 1.25, so over area is 2 pi times this = 7.85
%area out to 2.4 is pi 2.4^2 -- together we get an average of 0.43 
% integrating pressure over area we get 2*pi*1.25 rh^2 p_c which gives 2p_c L^2
It is convenient to write the velocity dependent aerodynamic damping force in terms of the acceleration
on the mechanism
\begin{equation}
a_{z,{\rm aero}} \sim \frac{F_{\rm aero}}{M} \approx - \alpha_z  \omega v_z  \label{eqn:a_z}
\end{equation}
with $\alpha_z$ a dimensionless parameter, 
\begin{equation}
\alpha_{z} \equiv \frac{\mu_{\rm air}}{\kappa } \frac{L^3}{8 M \omega}.  \label{eqn:alphaz} 
\end{equation}

In section \ref{sec:flow} we experimentally 
 estimate the coefficient $\alpha_z$, giving the force on 
 the mechanism due to air flow through the permeable substrate.
In section \ref{sec:ppflow} we modify equation \ref{eqn:a_z} to take into account
flow between mechanism base and  substrate when the mechanism is close to but
above the surface and using a Plane Poisseuille flow model for viscous flow between
two plates.
In section \ref{sec:model} we incorporate our estimates for $\alpha_z$ into
  numerical  models for the mechanism locomotion.  
  
\begin{figure}
\centering
\includegraphics[width=3.2in]{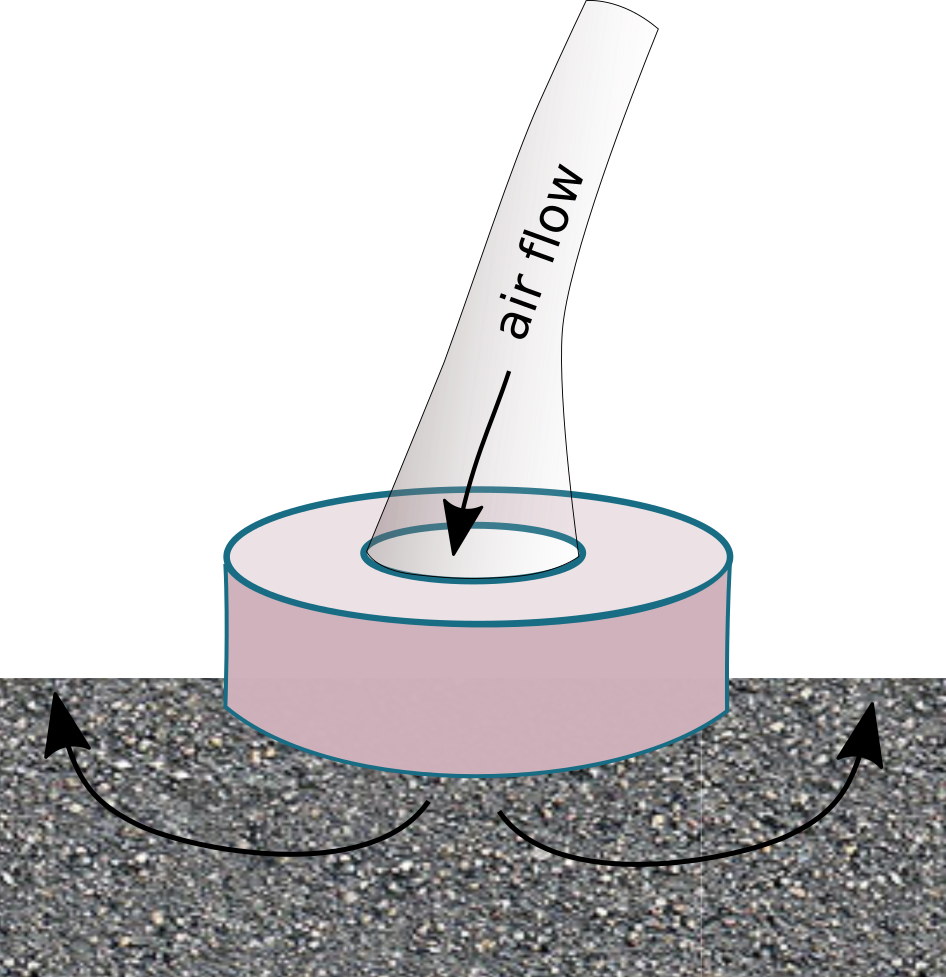} 
\caption{Flow rate vs pressure measurements were taken for air forced under a 
 flat annular block from its central hole while it was resting on a granular substrate.   
\label{fig:annular_block}}
\end{figure}
  
\begin{figure}
\centering
\includegraphics[width=3.2in]{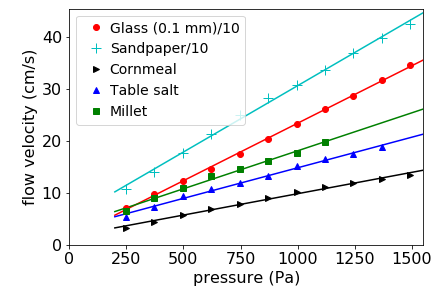} 
\caption{Flow velocity vs pressure measurements for air forced under a 
 flat annular block while it was resting on a substrate.   The points show measurements
and lines show linear fits to the data.  The line slopes are listed in Table \ref{tab:slopes}.
We use these slopes to estimate the suction from air pressure on the mechanism.
Five experiments are shown.  On a glass plate and on sand paper, we assume
that flow is horizontal.  The flow velocity, $u$, is calculated from the flow rate, $dV/dt$, using
the cross section area estimated in the middle of the annulus and depending on the gap width. 
For the block on cornmeal, salt and millet experiments, we assume that flow is through the granular medium
and the flow velocity is calculated from $dV/dt$  using the area of annulus' inside hole.
Because the velocities are higher on glass and sandpaper these velocities (for both points and lines)
have been divided by 10. 
\label{fig:flowpressure}}
\end{figure}

\subsection{Air flow rate vs pressure measurements}
\label{sec:flow}

To estimate the coefficient $\alpha_z$ in Eqn. (\ref{eqn:alphaz}), 
we experimentally measured 
how the air flow rate beneath a block and through our granular media depends on air pressure. 
We adopted a test geometry similar to that of our mechanism by placing a block  on
the surface of a granular substrate.
The block has a flat base that has an annular shape, with an outer diameter of $d_o = 6$ cm
and an inner hole with a diameter of $d_i = 2.5$ cm.    The air within the inner hole is placed under pressure
but outside the annular block the granular medium is open to atmospheric pressure, 
see Fig. \ref{fig:annular_block} for an illustration of the experiment.
A hose supplied
air  to the inner hole, and escaped through the substrate and the gap between the substrate and
the lower surface of the block.  

Air was supplied to the inner hole at controlled pressure (above atmospheric) using a bubble regulator.
Our device supplies air at  pressures of 250-- 2000 Pa  and uses water height to measure pressure.  
Pressure remained within  about $\pm 1/4$ in of water ($\pm 60$ Pa) of each set value.
%[4" of water ~= .01 atmosphere ~= 1000 Pa => 1" = 250 Pa]
We measured the air volume flow rate
using a home-made soap-bubble flow meter, where a soap film rises in a graduated
transparent tube of constant diameter.
Timing the transit between marks on the tube gave a measure of  the volume flow rate of air.
For each experiment we measured the flow rate at about 8 different set pressures.
%The internal diameter of the tube was 1.245" +- .005" (nice tube) = 3.1623 cm
%(square root of 10 to 5 significant figures... strange coincidence...)
During measurements, the annular block was weighted with a 12 oz weight (3.3 N) 
to prevent the air pressure from lifting the block and distorting the contact.
On the granular substrates, we ensured that the granular surface was flat by pressing and releasing the block before
air flow rates were measured.
%the block was preloaded and released sufficient to flatten the
%granular surface beneath the block, as happens quickly under a vibrating platform.
%Error bars on flow rates are a bit complicated. 
The percentage error in the pressure
at 1" H$_2$O is 25\% while the percentage error in the time is only a few percent.
At the higher end, the pressure measurement is good to 4\%, but the flow rates are only good to 10-15\%.
%Calculation is a division, so the percentages add.
%I've supplied raw data so we can compute individual error bars if we want.

We measured flow rates $dV/dt$ in cc/s for a block spaced 0.1 mm above a glass sheet,
on 120 grit sandpaper (115 $\mu$m particle sizes), on common table salt (with a mix of grain diameters 
in the range 0.25--0.5 mm), cornmeal and on millet.
Grain diameters for cornmeal and millet are listed in the notes to Table \ref{tab:quantities}.
For the experiments above a glass sheet  and on sandpaper, the substrates have solid bases
so air is restricted to travel in the narrow space between the block and the glass plate
or paper backing on the sandpaper.   For these two experiments
we estimate the air flow velocity $u =  \frac{dV}{dt} \frac{1}{a_w}$ in the middle 
of the block annulus using cross section area $a_w = \frac{2 \pi w (d_o + d_i)}{4}$. 
For the block on glass experiment the spacing between block and glass plate is $w = 0.1 $ mm.
For the  block on 120 grit sandpaper experiment we use a width $w = 115\ \mu$m, equal to the typical grain
size diameter for the grit on the sandpaper. 
The remaining experiments, on cornmeal, salt and millet, we assume that 
 the air flows down  through the granular medium.
We estimate the flow velocity $u = \frac{dV}{dt} \frac{1}{a_w}$  with area $a_w = \frac{\pi d_i^2}{4}$ computed 
with the block annulus' inside diameter.
The computed air flow velocity $u$ versus pressure $p$ measurements
are shown in Fig. \ref{fig:flowpressure}. 
We measured the  
slopes $S$ of each set of points  by fitting lines to the data points  and these slopes 
(in units of cm s$^{-1}$ Pa$^{-1}$)
are  listed in Table \ref{tab:slopes}.

\begin{table}
 \centering
\caption{Flow velocity vs pressure slopes}
\label{tab:slopes}
\begin{tabular}{lll}
\hline
%Slopes $K^{-1}$ are
Substrate & Slope $S$  \\ 
\hline
Glass plate, separation 0.1 mm &  0.22 \\
120 grit sandpaper               & 0.26 \\
Cornmeal                             & 0.0082  \\
Common table salt              & 0.012 \\
Millet                                   & 0.015  \\
\hline
\end{tabular}\\
Notes--   Slopes $S$ are given in  cm s$^{-1}$ Pa$^{-1}$.
These are the slopes of the lines shown in Fig. \ref{fig:flowpressure}.
%Flow velocities have been compute from measured flow rates ($dV/dt$) using
%a cross section area.  The area used for the granular media is the inside diamater of the annulus.
%For the glass plate and sandpaper the area used is that mid annulus and dependent on
%the width between annular block and surface.
We estimate $\pm 15\%$  uncertainty in the slope measurements.
\end{table}

The geometry of flow for our annular block is similar to that described for the air
flow under the mechanism described by Eqn. (\ref{eqn:uz}).
Fig. \ref{fig:flowpressure} shows  linear fits to the air flow velocity vs pressure measurements.
The nearly linear behavior supports our  approximation given in Eqn. (\ref{eqn:uz})
for air flow through a permeable medium caused by a pressure peak  on the surface.
A linear dependence of the flow rate on pressure is consistent with a low Reynolds number
regime where air viscosity and permeability of the substrate are important.
%We have checked that the coefficient measured on the glass plate is consistent
%with the prediction of Plane Poiseuille flow.
The lines in Fig.  \ref{fig:flowpressure} don't go through the origin, 
so we do see evidence
of non-linearity in the flow velocity vs pressure relation 
(for extensions to Darcy's law for airflow through agricultural grains see \cite{ergun52,smith96,molenda05}).  
%At low flow rates
 %the slope $S$   would be higher  than estimated here. 
However  a vertical mechanism platform velocity $A \omega = 40$ cm/s 
is well above the maximum measured flow velocity $\sim 18 $ cm/s measured on the granular media. 
To apply the flow rate vs pressure  measurements to our mechanism mechanics, 
we must  extrapolate to large values, rather than work in the low flow and  non-linear regime.

As was true in section \ref{sec:half} for air flow under the mechanism, we assume that the air flow
through a granular medium
can be described by Darcy's law.
The flow field below the substrate surface can again be approximated by Eqn. (\ref{eqn:prz})
and Eqn. (\ref{eqn:uz}) but using the outer radius of the block annulus instead of the half length 
of the mechanism as that is where the pressure must be equal to atmospheric pressure.
We take our experimentally measured pressure differential to be the central pressure 
$p_0$ in Eqn. (\ref{eqn:prz}) and
the measured air velocity to be the central velocity $u_0$  in Eqn. (\ref{eqn:uz}).
We expect the air flow velocity into the substrate inside the annulus 
\begin{align}
u \approx S \Delta p,  \label{eqn:af}
\end{align}
where $\Delta p$ is the air pressure differential and the slope $S$  for air velocity vs pressure must be
\begin{equation}
S \approx \frac{\kappa}{\mu_{\rm air}} \frac{2.4}{d_o/2},  \label{eqn:S}
\end{equation}
following Eqn. (\ref{eqn:u0}).
Using the viscosity of air (listed in Table \ref{tab:quantities}) and
the slopes we measured on cornmeal, salt and millet (and listed in Table \ref{tab:slopes})  Eqn. (\ref{eqn:S})
gives permeabilities of $\kappa = 1 $ -- $ 3 \times 10^{-7}\ {\rm cm}^{2}$.
For loose sand the permeability ranges from $\kappa \sim 10^{-5} $ to $10^{-8}\ {\rm cm}^{2}$.
Our measured slopes are consistent with permeability measurements in porous media.

A comparison of Eqn. (\ref{eqn:af}) and (\ref{eqn:S}) with Eqn. (\ref{eqn:uz}) implies that we can estimate
the dimensionless coefficient $\alpha_z$ (Eqn. \ref{eqn:alphaz}) with  slopes $S$ measured here
and correcting for the ratio of the mechanism  length  and the block annulus' outer diameter,
\begin{align}
\alpha_z &= \frac{1}{2 S} \frac{L}{d_o}  \frac{L^2}{M \omega} \\
&= 2 \ \left( \frac{0.015\ {\rm cm\ s^{-1}\ Pa^{-1}}}{S}\right)  
	\left( \frac{L}{4\ {\rm cm}}\right)^3
	\left( \frac{1.4\ {\rm g}}{M} \right) \nonumber \\
&~ ~ ~ ~  ~ ~ ~ ~ \times
	\left( \frac{f}{12,000\ {\rm rpm}}\right)
	\left( \frac{6\ {\rm cm}}{d_o }\right).  \label{eqn:alphaz_meas}
\end{align}
The dimensionless parameter characterizing the suction $\alpha_z \gtrsim 1$, 
supporting our hypothesis that air pressure can affect the mechanism's
dynamics.  While we have described the force as a suction force, it would also operate to cushion
the mechanism when it lands on the substrate.

\subsection{Plane Poiseuille flow}
\label{sec:ppflow}

Plane Poiseuille flow describes steady laminar flow in a viscous fluid between two parallel sheets separated by
a narrow distance $z$.  Plane Poiseuille flow also obeys a relation between pressure
gradient and flow velocity, similar to Darcy's law (Eqn. \ref{eqn:Darcy}) and similar to our measured
relation between flow velocity and pressure in Eqn. (\ref{eqn:af}).   
For Plane Poiseuille flow,
a mean flow speed $ u$  (averaged over the velocity profile between the plates) obeys
\begin{equation}
 u = -\frac{z^2}{12\mu_{\rm air}} \nabla p.    \label{eqn:PP}
\end{equation}
Here air flow is parallel to the plates as there is no flow through their surfaces.
The similarity between Eqn. (\ref{eqn:Darcy}) and  Eqn. (\ref{eqn:PP}), relating flow speed to a pressure gradient,
implies that Plane Poiseuille flow is consistent with an effective 
permeability $\kappa_{\rm PP}(z) = z^2/12$ that depends on distance between the plates, $z$.

We compare our flow velocity vs pressure measurements to the predictions of Plane Poiseuille flow
and then we will modify our estimate for the vertical drag force due to air pressure to take
into account air motion parallel to the mechanism base when the mechanism is near but not on the 
granular substrate.

In Table \ref{tab:slopes} and with measurements shown 
Fig. \ref{fig:flowpressure} we measured a flow velocity vs pressure for the annular block separated by 0.1 mm
from a glass plate.  We estimate the pressure gradient $\nabla p = 2\Delta p/(d_o - d_i)$ 
across the annulus.  This and Eqn. (\ref{eqn:PP})
gives a predicted slope (describing pressure vs flow velocity) of  
$S = \frac{z^2}{12 \mu_{air}} \frac{2}{d_o - d_i}  $. 
With $z=0.1 $ mm, this gives $S = 0.27$ cm~s$^{-1}$~Pa$^{-1}$, and is consistent
with the 0.22 cm~s$^{-1}$~Pa$^{-1}$ slope we measured.
Our pressure vs flow rate measurement for the block near a glass plate  are consistent with 
that estimated for Plane Poiseuille flow.

Taking into account %the Bernoulli effect (Eqn. \ref{eqn:Bernoulli}) or 
viscous Plane Poiseuille flow would
give a height dependent aerodynamic acceleration on our mechanism $a_{z,{\rm aero}}(z)$.  
We use Eqn. (\ref{eqn:dVdt}) to relate horizontal air speed $u$ to vertical platform velocity $v_z$, 
Eqn. \ref{eqn:PP} for Plane Poiseuille flow, estimate the pressure gradient under the mechanism
as $\nabla p \sim 2p_0/L$ with $p_0$ under the platform and the force on the mechanism as $p_0 L^2$.
This gives an estimate for the 
 aerodynamic force on the mechanism  due to air
flow beneath the mechanism
\begin{align}
F_{\rm PP,aero} \sim -\frac{\mu_{\rm air}}{z^2/12} \frac{L}{z} \frac{L^3}{8  } v_z.
\end{align}
This equation resembles the aerodynamic force we estimated from permeability alone; Eqn. (\ref{eqn:Faero}), except
the force becomes large as $z\to 0$.  
We expect the force cannot be higher than that estimated for the permeable medium.
We can modify  the acceleration  
of Eqn. (\ref{eqn:a_z}) to make a 
height dependent transition   
\begin{align}
a_{z,{\rm aero}}(z) = -\alpha_z v_z \omega \left( \frac{1}{1 +  (z/h_{PP})^3} \right), \label{eqn:z3}
\end{align}
and retaining the definition of the dimensionless parameter $\alpha_z$ of Eqn. (\ref{eqn:alphaz}).
The length that sets the transition between regimes
\begin{align}
h_{PP} &= \left( 12 L \kappa  \right)^\frac{1}{3} \nonumber \\
&= 0.17 \ {\rm mm} \left(\frac{L}{4\ {\rm cm}} \right)^\frac{1}{3}  \left( \frac{\kappa}{10^{-7} {\rm cm}^2}\right)^\frac{1}{3} 
\end{align}
and where we used our estimate for the permeability $\kappa$ from in section \ref{sec:flow} 
(see Eqn. \ref{eqn:S}).
We define a dimensionless length $\bar h_{PP}$,
\begin{align}
\bar h_{PP} &= \frac{h_{PP}}{A} \approx 0.5 \left(\frac{L}{4\ {\rm cm}}  \right)^{\frac{1}{3}}
\left( \frac{\kappa}{10^{-7} {\rm cm}^2}\right)^\frac{1}{3}
\left( \frac{0.3\ {\rm mm} }{A} \right). \label{eqn:barhPP}
\end{align}

\section{Numerical Model for Locomotion}
\label{sec:model}

We describe the mechanism motion in  two dimensions   with $x,z$ 
corresponding to horizontal and vertical coordinates.   We use $\bar x = x/A $ and $ \bar z = z/A$  for 
the coordinates in units of  the vibration displacement amplitude $A$.
Time $\tau$ is
in units of $\omega^{-1}$ with $\tau = t \omega$.     Velocity is in units of $A \omega$
and acceleration in units of $A\omega^2$.
We assume that the mechanism
base remains parallel to the granular substrate, with $\bar z$  giving the distance between
substrate and mechanism base (in units of $A$)  and $\bar z =0$ for platform base touching a 
flat substrate.  The center of the mechanism has $\bar x=0$.
We ignore tilting, rocking, flexing and turning.

The equation of motions for our model resembles of that by \cite{jalili16} for harmonically driven
micro-robots that either hop or move via stick-slip friction interactions.
Our equation of motion in $\bar z$  for the platform base center is 
\begin{equation}
\frac{d^2 \bar z}{d \tau^2} = -\Gamma^{-1} + \cos (\tau + \phi_0) + a_{z,{\rm aero}}
\label{eqn:ddotz}
\end{equation}
where $\phi_0$ is an initial phase for the motor. 
The recoil from the internal motion of the motor flywheel gives a sinusoidal acceleration
of amplitude 1. 
The constant term, inversely dependent on the acceleration parameter, is due to 
 gravity.    The rightmost term is from the force due to aerodynamics.  
The acceleration due to air pressure should only be significant when the mechanism is nearly touching the substrate.
To reduce the aerodynamic acceleration as a function of height we assume
\begin{equation}
a_{z,{\rm aero}} = 
  -\frac{d {\bar z}}{d\tau} \alpha_z \exp(- \bar z/h_m). \label{eqn:az}
\end{equation}
The coefficient $\alpha_z$ is estimated in section \ref{sec:aero}.
Because the pressure should drop as the mechanism moves upward, we cut off
the areodynamic force with a distance $h_m$ that we treat as a free parameter.
We expect $h_m \sim \bar h_{PP}$ estimated in Eqn (\ref{eqn:barhPP}) from the transition
to Plane Poiseuille flow.  
The Bernoulli effect estimate for pressure could also contribute
to aerodynamic force decreasing with mechanism height.  Because low Reynolds number
flow is a poorer approximation at larger $\bar z$, we cut off the friction exponentially
rather than with a power law (as in equation \ref{eqn:z3}).  The model gives
similar trajectories with a ${\bar z}^{-3}$ cutoff in the drag force.

Our adopted equation of motion for $\bar x$ is similar to Eqn. (\ref{eqn:ddotz}) but does not depend on
 gravity,
\begin{equation}
\frac{d^2 \bar x}{d \tau^2} =      \sin (\tau  + \phi_0) + a_{x,{\rm drag}} 	\label{eqn:x}
\end{equation}
with an additional  height dependent horizontal drag  force  $a_{x,{\rm drag}}$.

We assume that the mechanism does not bounce off the substrate.
When the trajectory reaches $\bar z=0$, with $\frac{d\bar z}{d\tau} <0$, and contacts the surface,
we set the vertical velocity component to zero,
$\frac{d\bar z}{d\tau} =0$.  
At the same time in the integration, we use a Coulomb friction law to set the 
horizontal component of acceleration due to contact with the surface.
The frictional horizontal acceleration depends on the normal force exerted by the surface.
With $v_{x0}, v_{z0}$  the velocity components prior to impact
we set the horizontal velocity after impact to be 
\begin{equation}
\frac{d \bar x}{d \tau}   = 
\begin{cases}
  v_{x0} - \text{sign} (v_{x0}) \mu_s |v_{z0}| &  \text{for} \ \   |v_{x0}| > \mu_s | v_{z0}| \\
  0 & \text{otherwise} 
\end{cases}
  \label{eqn:du}
\end{equation}
with $\mu_s$ the coefficient of friction which we set to the static value even though
friction occurs when the mechanism is sliding on the substrate.

We allow a small horizontal drag force to be present, that is 
 in the same form as  (Eqn. \ref{eqn:az})
\begin{equation}
a_{x,{\rm drag}}(\bar z) = -  \alpha_x \frac{dx}{d\tau} \exp(- \bar z/h_m),  \label{eqn:ax}
\end{equation}
 but with a different
coefficient $\alpha_x \ne \alpha_z$.   
Our high speed videos illustrate that
grains are occasionally pushed and levitated as the mechanism moves.
These interactions would slow the mechanism, acting like a horizontal drag force.
Locomotion studies in granular media have previously adopted a hydrodynamic-like 
velocity dependent drag force
(e.g., \cite{katsuragi13,aguilar16b}), though in these settings the locomotor is massive
enough to penetrate into the medium.  Even though  the air pressure
affects the vertical acceleration,  shear in the air flow cannot  exert a significant traction force horizontally.
However if there is some contact with the surface when suction is strong, then the normal force
on the mechanism could be higher than computed using
gravitational and motor recoil normal forces alone and this would increase the friction force.
%A variant of equation \ref{eqn:ax} would be a Coulomb friction term that takes into a 
%normal force that is enhanced by suction.
 
A challenge of numerically integrating a damped bouncing system is the 
 ``Zeno'' effect, in which the number of bounces can be infinite in a finite length of time.
Also because our forces depend on height, it is not straightforward to 
 integrate from bounce to bounce, as commonly done for modeling a ball bouncing
on a vibrating table top \cite{holmes82}.  In our integrations  we take short time-steps 
and check for proximity
to ${\bar z}=0$ each step.
We update positions and velocities using straightforward first order finite differences. 
We chose the time step to be sufficiently small that the 
the trajectories are not dependent on it ($d\tau = 0.01$).

\begin{figure*}
\centering
$\begin{array}{c}
\includegraphics[width=4.7in, trim={0mm 2mm 0mm 2mm},clip]{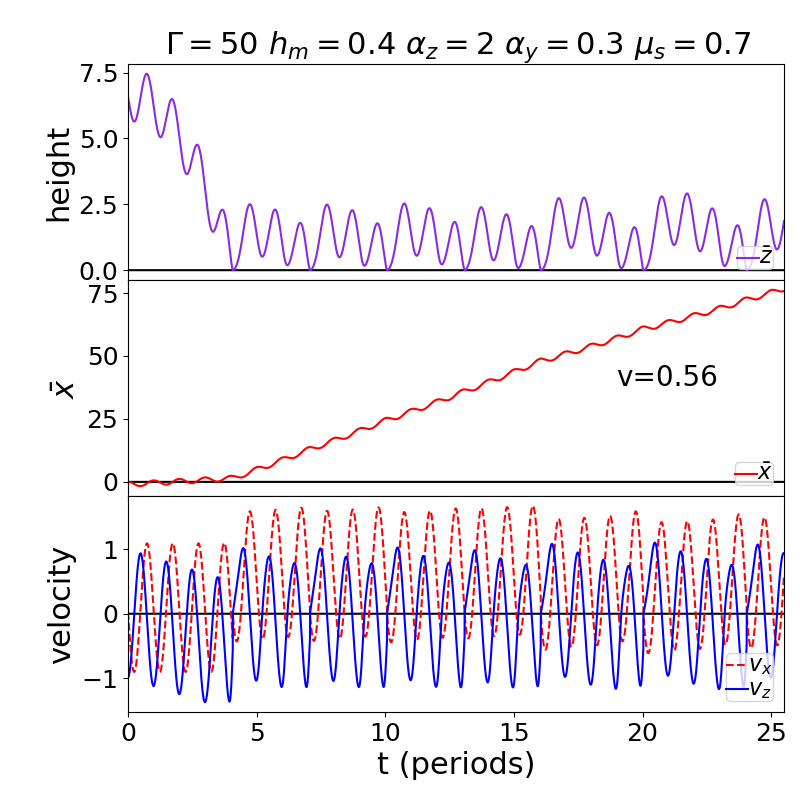} \\
\includegraphics[width=6.0in, trim={-23mm 70mm 0mm 90mm},clip]{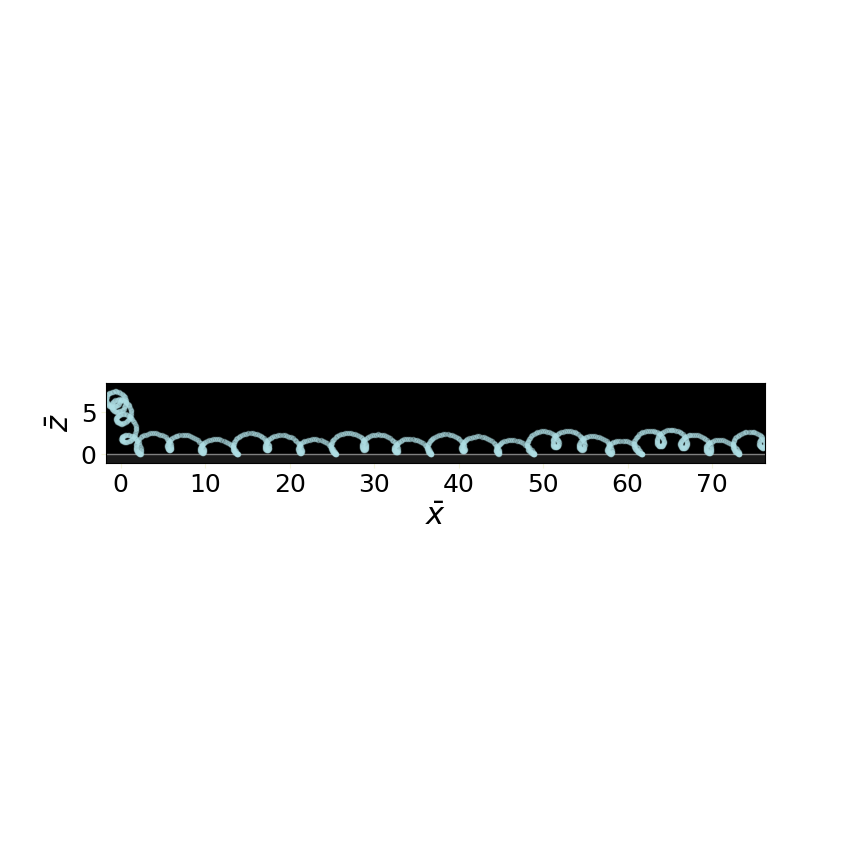} 
\end{array}$
\caption{a) An integrated numerical model for mechanism locomotion. 
The top panel shows vertical motion or $\bar z(\tau)$, the middle plane shows horizontal motion or
$\bar x(\tau)$.  The bottom panel shows  velocity components as a function of time.
The trajectory is integrated for 25 motor oscillation periods using Eqn. (\ref{eqn:ddotz}) and Eqn. (\ref{eqn:x}).
In the middle panel the mean horizontal speed is labelled on the plot.  The speed 
was computed with a linear fit to $d{\bf x}/d\tau$ at times $\tau > 10$.
Coefficients for the model are shown above the top panel with $\Gamma$ the acceleration
parameter, $\alpha_z$ and $\alpha_x$ setting the vertical and horizontal damping forces,
the height $h_m$ setting the cutoff height for them and $\mu_s$ the coefficient of friction 
for the substrate.
Time is in units of motor angular rotation frequency, $\omega^{-1}$, distances are in
units of mechanism vibration amplitude  $A$, and velocities are in units of $A \omega$.
b) The model mechanism trajectory $\bar x$ vs $\bar z$ is shown for the same numerical integration
and can be compared to photographed trajectories in Figs. \ref{fig:wiggle} and \ref{fig:doubling}.
\label{fig:traj3}}
\end{figure*}

\begin{figure*}
\centering
\includegraphics[width=6.0in, trim={0mm 20mm 0mm 10mm},clip]{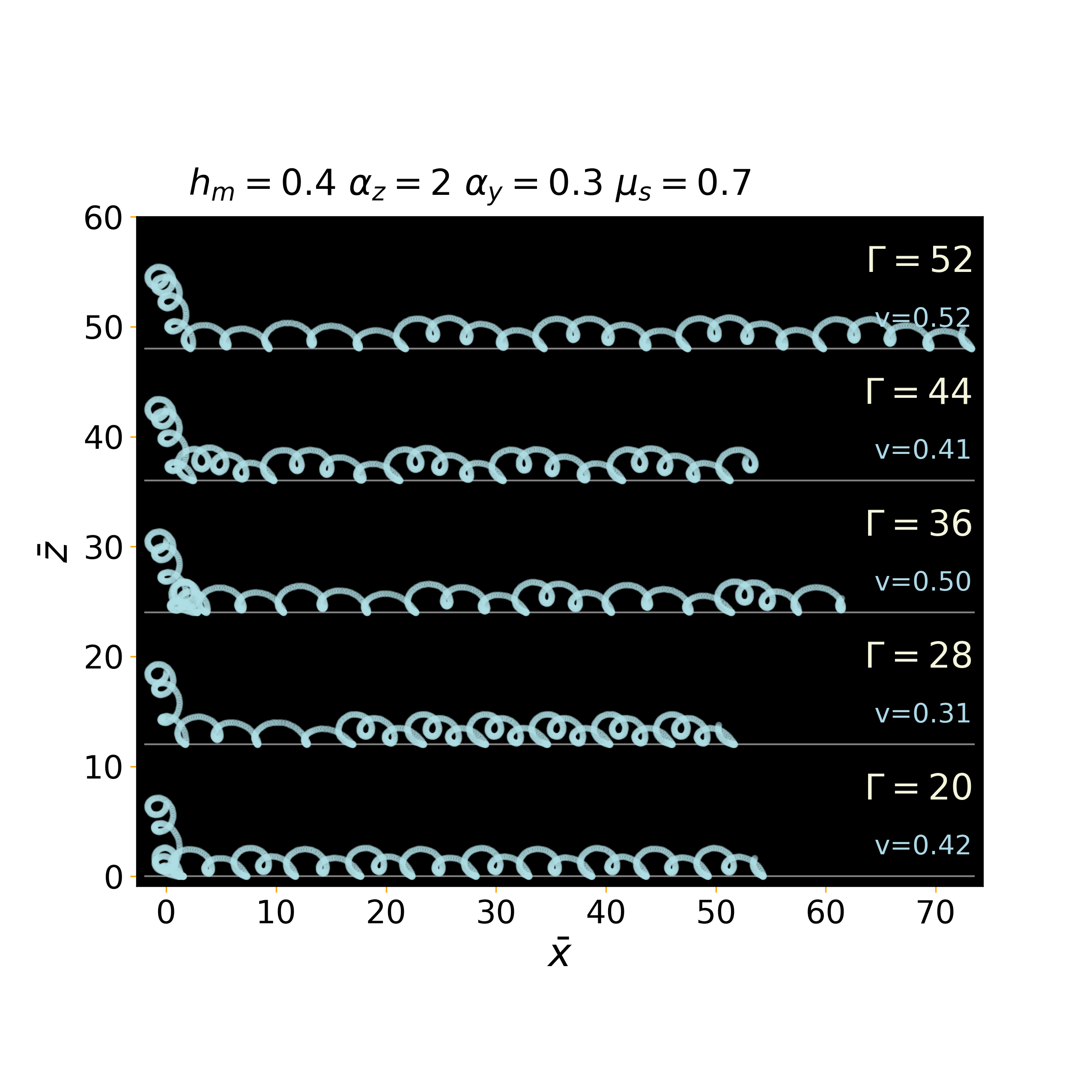} 
\caption{Model mechanism trajectories for different acceleration parameters, $\Gamma$.
The higher acceleration parameter ones correspond to
mechanisms with higher frequency motors.   Average horizontal speeds  are labelled
for each trajectory on the right.
Even though velocity in units
of $A \omega$ are similar for these trajectories,  the higher $\Gamma$ 
models correspond to faster actual mean horizontal speeds  
as our unit of velocity is not independent of $\Gamma$.
\label{fig:v_gam}}
\end{figure*}

\begin{figure*}
\centering
\includegraphics[width=6.0in, trim={0mm 20mm 0mm 10mm},clip]{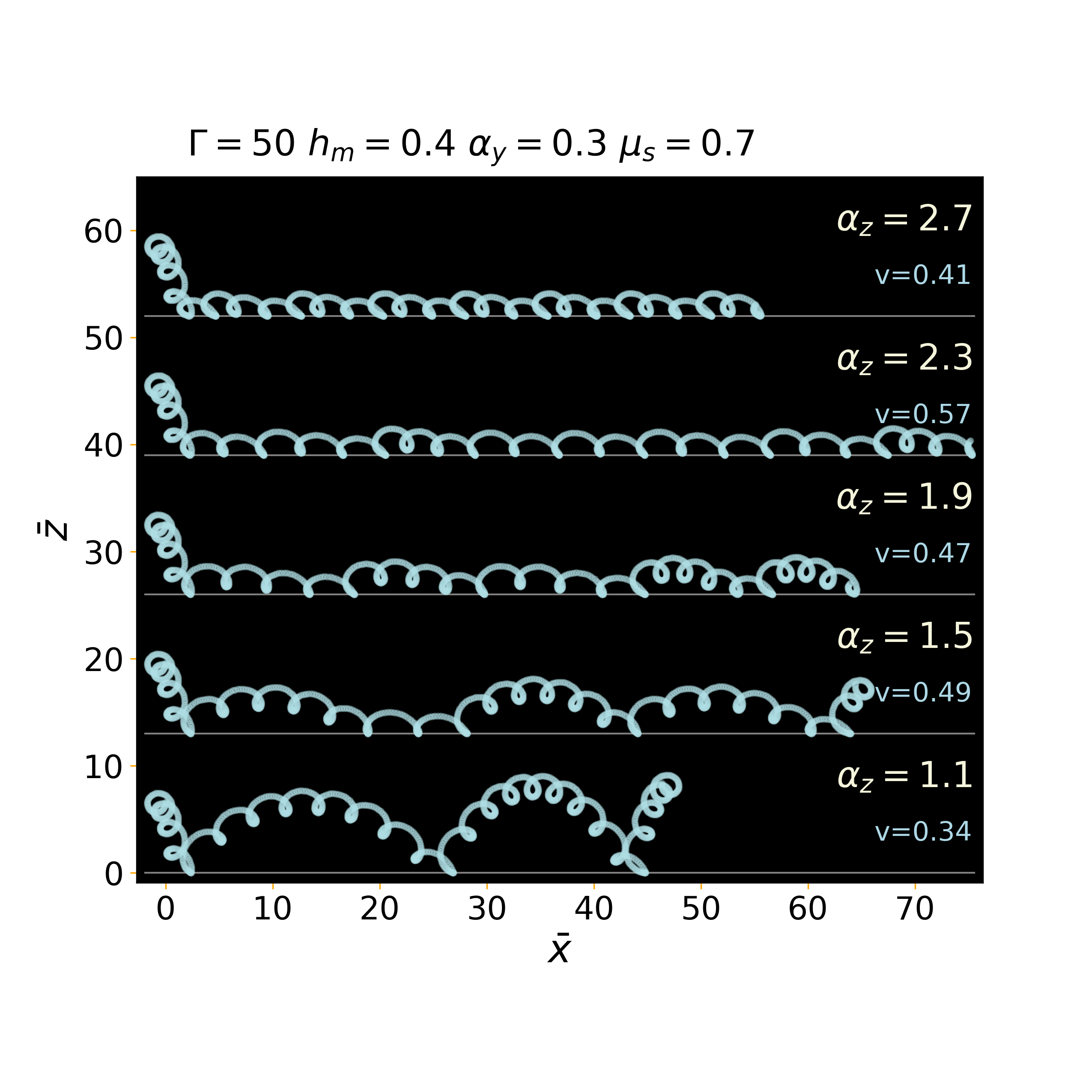} 
\caption{Model mechanism trajectories for different damping parameters, $\alpha_z$.
The fastest mechanism is  second from top at $\alpha_z = 2.3$.
This is faster because it gallops more efficiently, pushing off each period, instead of
hopping into the air and pushing off the ground every few periods as is true of the lowest trajectory.
The topmost trajectory is slower because the suction is so high the mechanism does not launch
itself off the ground for very long during each motor oscillation.
\label{fig:v_az}}
\end{figure*}

\subsection{Model mechanism trajectories}
\label{sec:traj}

Fig. \ref{fig:traj3} shows a trajectory computed using Eqn. (\ref{eqn:ddotz}) -- Eqn. (\ref{eqn:ax}).
Parameters  for the model, $\Gamma, \alpha_z, \alpha_x, h_m$ and $\mu_s$, are listed above the top panel.
Our dynamical system is similar to a vertically bouncing ball on a vibrating sheet
(e.g., \cite{holmes82,germinard03,pieranski83,bapat86}).  
The bouncing ball can be described with a map between
bounces, during which time the ball trajectory is specified by gravity alone \cite{holmes82}.
Each bounce instantaneously changes the ball's vertical velocity.
This gives an initial condition that can be used to determine when the next bounce occurs.
The dynamical system is rich, exhibiting period doubling and chaos 
\cite{holmes82,pieranski83,pieranski85}.
The horizontal motion
gives an additional degree of freedom that can increase the complexity of the problem.
For the bouncing ball above a vertically oscillating but parabolic plate, the vertical and horizontal motions
are  fully coupled; even a small curvature in the plate can induce chaotic behavior
 \cite{mcbennet16}.  Here our $x$ equation of motion depends on $z$ but the $z$ equation
 of motion does not depend on $x$.  
 
 In Fig. \ref{fig:traj3} 
the mechanism is begun above the substrate and initially is in free fall. While in free fall, the base of the mechanism
oscillates due to the motor recoil.  At $\tau \sim 5$, it contacts the surface, and
 its vertical velocity is reduced to zero. This happens each time it contacts the surface. 
Afterward touching the substrate, the velocity rises slowly because the acceleration 
must overcome the vertical drag force.
The phase of oscillation is such that the horizontal velocity is near its minimum value when the motor is near
the surface.  Friction with the substrate and the horizontal drag force bring this minimum toward
zero. Because of the motor recoil, the mean horizontal velocity of the mechanism is above its minimum, 
and this gives the mechanism a net forward horizontal motion.

After integrating the equations of motion, the trajectory is resampled so that points are equally spaced
with respect to time. In Fig. \ref{fig:traj3}b the trajectory is plotted with partially transparent points 
giving a lighter line where the velocity is slower.    
The model trajectory mimics the LED brightness traced  in the photographs shown
in  Fig. \ref{fig:wiggle}.
For the motor at 5.5V at 285Hz, the acceleration parameter is about 50
and the velocity of motion caused by recoil is about $A \omega = 52$ cm/s.   The velocity of
the mechanism  we measured to be about 30 cm/s on poppy seeds, which is 0.57 in 
units of $A\omega$.    We set $\alpha_z= 2$ with Eqn. (\ref{eqn:alphaz_meas})
 consistent with $f=285$ Hz,   
 and using the slope of $S = 0.012$ cm~s$^{-1}$~Pa$^{-1}$
measured on salt which has a similar grain size as poppy seeds.
To match the vertical height exhibited by the trajectory we adjusted $h_m$
and we adjusted $\alpha_y$  to match the horizontal speed.    We did not adjust the substrate coefficient
of friction, setting it equal to the coefficient of static friction, $\mu_s$, for our granular media.  
The value of $h_m = 0.4$, giving realistic looking model trajectories, is consistent with 
that estimated for $\bar h_{PP}$, the height setting the transition 
 between  Plane Poiseuille and permeable flow (see Eqn. \ref{eqn:barhPP}).  

In Fig. \ref{fig:v_gam} we show the effect of varying acceleration parameter
and in Fig. \ref{fig:v_az} the effect of varying the strength of the aerodynamic force.
Period doubling is seen in Figs. \ref{fig:traj3}b,   \ref{fig:v_gam} and \ref{fig:v_az}, similar to that seen in the
the photographs (Fig. \ref{fig:wiggle} and Fig. \ref{fig:doubling}).
Some of the non-linear phenomena exhibited by the bouncing ball on the vibrating
table top \cite{pieranski83} is also exhibited by our simple  vibrating mechanism model.

Figure \ref{fig:v_gam} shows that trajectory speeds in units of $A\omega$ are not strongly
dependent on acceleration parameter.  However actual horizontal speeds at larger acceleration parameter
 $\Gamma$ are likely to be 
higher as the velocities are in units of $A\omega$.   Mechanisms with larger vibrational amplitudes
(larger recoil) and faster vibration would move horizontally faster.

While  the model trajectories are not strongly dependent
on acceleration parameter,  they are, as seen in Fig. \ref{fig:v_az},  dependent
on the strength of the vertical drag force.   We have noticed that mechanism motion is often jumpier on
rougher substrates.   The parameter $\alpha_z \propto S^{-1}$ (Eqn. \ref{eqn:alphaz_meas}) is smaller
on rougher substrates, as we saw  our pressure vs flow measurements $S$ is larger for millet
than salt or cornmeal.  This is
consistent with smaller $\alpha_z$
models having trajectories that reach larger heights.
Similar trajectories are observed without any horizontal damping force, $\alpha_x =0$,
 though the model is more chaotic, exhibiting speed variations 
 and the mechanism sometimes changes direction entirely.

Horizontal speeds are labelled on each panel in Fig. \ref{fig:v_az}
and show that
the faster mechanism is  the second from top with $\alpha_z = 2.3$.  Also the length of
the trajectories in this figure is related to the average horizontal speed as each trajectory was integrated
for the same time length of 25 motor periods.
The fastest trajectory is fast because it gallops more efficiently, pushing off each period, instead of
hopping into the air and pushing off the ground every few periods as is true of the lower trajectories.
The topmost trajectory is slower because the vertical drag  is so high the mechanism does not launch
itself effectively off the ground.  This figure illustrates a speed optimization strategy for
design.  Mechanisms that are designed so that their center of mass does not make 
large vertical excursions during locomotion would be faster than those that propel
the mechanism   to larger heights, as  during large vertical jumps the mechanism cannot  push itself forward
 by kicking the surface.
 
When the motor frequency 
is changed, both acceleration parameter and estimated $\alpha_z$ vary, 
as our estimate for $\alpha_z$ is inversely proportional to motor oscillation frequency (see
Eqn. \ref{eqn:alphaz_meas}).   At slower motor frequencies the acceleration parameter
$\Gamma$ is lower and the vertical damping parameter $\alpha_z$ is higher. 
Figure \ref{fig:wiggle} shows that at slower motor frequencies
the trajectories are more irregular, and opposite to that predicted by the model as higher $\alpha_z$
models tend to have flatter trajectories.
The high frequency trajectories on cornmeal, poppy seeds and millet shown in Fig. \ref{fig:wiggle} are similar,
however their permeabilities differ and with different $\alpha_z$ the modeled trajectories
would have different shapes.
We have noticed that the numerical model predicts similar trajectories 
for different $\alpha_z$ but  at fixed $\alpha_z h_m$ and $\alpha_z/\alpha_y$.
The parameters describing our numerical model are not  independent.
Perhaps the effective cutoff height parameter $h_m$ should also be chosen to depend on the substrate.
This is not unreasonable as finer grained materials have higher $\alpha_z$ and the high air pressure
should primarily be important very close to the surface.  

While our simple numerical model is successful at reproducing the shape of mechanism trajectories,
and the model is based on the size of the vertical drag force  from flow vs pressure measurements,
the description of the aerodynamic and friction forces needs improvement to be more predictive.
Our model also neglects mechanism tilt, surface irregularities, flexing in the mechanism itself
and waves traveling along its power wires.  An improved model could include these degrees
of freedom in the model.

\begin{table}
 \centering
\caption{Additional nomenclature}
\label{tab:nomenclature}
\begin{tabular}{lll}
\hline
%Slopes $K^{-1}$ are
acceleration parameter & $\Gamma$ \\
pressure & $p$  \\ 
density & $\rho$ \\
permeability & $\kappa$ \\
Bessel function & $J_0()$ \\
cylindrical coordinates & $r, z$ \\
time   & $t$ \\
inside diameter & $d_i$ \\
outside diameter & $d_o$ \\
flow velocity vs pressure slope & $S$ \\
air flow rate & $dV/dt$ \\
air velocity & ${\bf u}$\\
mechanism velocity & $ {\bf v}$ \\
force & $F$ \\
acceleration & $a$ \\
initial motor phase & $\phi_0$ \\
Cartesian coordinates for model & $\bar x, \bar z$ \\
time coordinate for model  & $\tau $ \\
model parameters & $ \alpha_z, \alpha_x, h_m$ \\
\hline
\end{tabular}\\
%Notes--   %slopes are given in Pa s /cc.
\end{table}

\section{Summary and Discussion}
\label{sec:sum}

We have constructed a limbless, small (4 cm long), light-weight (less than 2 g) and low cost (a few dollars)
mechanism, similar to a bristle bot, but with a coin vibrational motor on a light foam
platform rather than bristles.   The mechanism traverses
granular media at speeds of up to 30 cm/s or 5 body lengths per second.
In units of body lengths per second our mechanism speed is similar to the 
the six legged  DynaRoACH  robot (10 cm long, 25 g)   \cite{zhang13}, but
slower than the  zebra-tailed lizard  (10 cm long, 10g)
that can move 10 body lengths/s \cite{li12}.   Our mechanism horizontal speed exceeds many
 conventional bristle bots, has no external moving parts and can traverse flat granular media.

We estimate the mechanism's 
vibrational acceleration from the motor recoil  divided by gravitation acceleration.
Our mechanisms can have acceleration parameters as large as 50.
They would be classified as a hopper or galloper in terms of their gait or walking Froude number.

With an LED mounted to the mechanism and with long exposures, we photographed 
mechanism trajectories during locomotion.  The mechanism trajectories are typically periodic, touching
the granular substrate once per motor period,
but sometimes they show period doubling or tripling, where the mechanism touches the substrate
 once every two or three motor oscillation periods.
The large acceleration parameters imply
that the trajectories should jump higher off the surface than they do when they are
undergoing periodic motion.   We infer that there must be a downward vertical force
that keeps the mechanisms close to the surface.    Following experimentation of mechanisms
on different surfaces and with different bases, in vacuum and on granular media covered in powder, 
and with high speed videos we
conclude that aerodynamics affects their locomotion.  

Using experimental measurements of air flow rate  vs pressure under blocks placed on different media, we estimated
the size of a vertical aerodynamic  force that is a suction force when the mechanism leaves
the surface.  The aerodynamics is modeled
using Darcy's law for flow through permeable media and Plane Poisseuille flow.
In both settings air flow velocity is proportional to pressure gradient due to low Reynolds number
flow in narrow spaces.
When incorporated into a numerical model the additional
force lets us
 match mechanism trajectory shapes and speeds.  The model is better behaved
 with a small additional horizontal drag force, that might arise as some grains are disturbed
 and the power wires can transmit momentum.
The model illustrates that speed is optimized by having large vibration amplitudes,
large vibration frequencies and a periodic trajectory that  touches the surface once
per oscillation period.   Our mechanisms may be self-optimized if the mechanism
platform flattens the substrate sufficiently to give effective suction as it traverses the medium.
In the absence of air (for example on asteroid regolith or Martian sand-dunes)
a design could  optimize speed 
by allowing the mechanism itself to flex.   For example a moving tail could act to limit
the vertical motion and optimize speed this way, similar to the way the counter motion of a kangaroo's tail reduces
the height a kangaroo reaches during each jump and minimizes the up and down motion 
of its center of mass.

The sensitivity of our kinematic model to air viscosity and substrate permeability  suggests that
construction and 
optimization of wider, lighter or smaller mechanisms will depend on these parameters.  
Larger mechanisms that can traverse granular media with similar dynamics to our mechanisms
might be constructed by designing the mechanism so that $\alpha_z \sim 1$, following Eqn. (\ref{eqn:alphaz}).
Despite their sensitivity to aerodynamics, we have found that the few gram hopper mechanisms 
we have constructed are fairly robust, 
can traverse solid surfaces as well as a variety of granular medium, can traverse granular media in a vacuum,
and work when constructed from different types of light platform materials.

Our mechanisms are not 
autonomous or maneuverable.  Additional capabilities would add
weight to the mechanism and as the mechanism
speeds depends on the mechanism recoil amplitude, this would reduce its speed.   
Light weight motors are available with larger recoil, but they tend to be cylindrical form 
with recoil motion vector traversing a cone rather than coin form with recoil confined to plane (as used here). 
The cylindrical  motors would be more complex to model.
A BEAM-robotics style \cite{beam} 
%\footnote{\url{https://en.wikipedia.org/wiki/BEAM_robotics}} 
autonomous locomotor might be achieved by constructing the mechanism platform from a light-weight solar panel.
As our mechanisms are inexpensive, large numbers of autonomous BEAM-robotics style locomotors
could be used for distributed exploration problems.

\begin{acknowledgment}
%acknowledgements.
ACQ is grateful to the Leibniz-Institut f\"ur Astrophysik Potsdam for their
warm welcome and hospitality July 2017 and April 2018, Mt Stromlo Observatory
for their hospitality winter 2018, 
and support from the Simons Foundation.

This material is based upon work supported in part by the University of Rochester
and National Science Foundation
Grant No. PHY-1460352 to the University of Rochester Department of Physics and Astronomy's REU program.

We thank Scott Seidman, Michiko Feehan,  Eric Nolting and  Mike Culver for help in the lab. 
%Thank  if not a co-author.
%and Jonathan Carroll-Nellenback, 
%Mark (Rosemary's SO) and for helpful discussions.
\end{acknowledgment}

\bibliographystyle{asmems4}
\bibliography{refs}

\end{document}